# Odds for the Brazilian 2018 president elections:
# An application of Bayesian statistics in contingency tables

Carlos Alberto de Bragança Pereira[1], Teresa Cristina Martins Dias[2], Adriano Polpo[2]
[1] *University of Sao Paulo and University of Mato Grosso do Sul, Brazil, cpereira@ime.usp.br*
[2] *Department of Statistics, Federal University of Sao Carlos, Brazil, dtmd@ufscar.br, polpo@ufscar.br*

The purpose of these notes is to present an assessment of the probability of a candidate be elected in a two-round presidential election. In the first round, all candidates can be voted on. If one of them has more than 50% of the vote (s)he is elected and there is no second round. If none of the candidates obtain more than 50% of the votes, then the top two candidates will be selected for a second round. In this second round, the most voted candidate is elected. This is the scenario of the Brazilian elections that are taking place at the moment. We are calculating the odds associated with the 2018 presidential elections in Brazil. The first round is on October 7, and the second round is on October 28, 2018.

There are thirteen candidates in the present president elections in Brazil. These candidates (political party), in alphabetic order, are: 1. Alvaro Dias (Podemos), 2. Cabo Daciolo (Patriota), 3. Ciro Gomes (PDT), 4. Eymael (DC), 5. Fernando Haddad (PT), 6. Geraldo Alckmin (PSDB), 7. Guilerme Boulos (PSOL), 8. Henrique Meirelles (MDB), 9. Jair Bolsonaro (PSL), 10. Joao Amoedo (Novo), 11. Joao Goulart Filho (PPL), 12. Marina Silva (Rede), 13. Vera Lucia (PSTU).

Considering a Bayesian framework, the multinomial model and the conjugate Dirichlet-family, we present the calculus of the probability of a candidate be elected. The data used here are public and are from the two most known companies, IBOPE and Datafolha. For Datafolha, we have four data sets, and for IBOPE, we have six data sets. All data sets are the most recent polls report of each company. Non-informative priors are used for the first poll of each company. For the subsequent polls, we use a scaled prior based on the posterior of the previous poll. The data are available at Poder 360 (2018). The number of interviews (sample size) and the percentage of each candidate for the first round are the data reported by each company. For the second round, the percentages of votes for each candidate are obtained for some scenarios with two of the candidates. The companies also reported, for each round, the percentage of undecided voters. We must call the attention of the readers that we are only responsible for the methodology of calculus and not for the sampling interviews framework. Data used here are from public sources and the authors have no relation of any kind with the two companies.

We are assuming, by using the multinomial model, simple random samples, a standard constraint for our model. The only objective of this work is to illustrate how to evaluate probabilities of interest. We do not discuss the quality of candidates or whether a name can be elected. Using sample percentages and sample sizes, one obtains the sample frequencies required for the calculation work. Thus for each research performed we have the vector $(x_1, \ldots, x_{15})$ of the frequencies voters for all candidates, blank votes and undecided. $n' = \sum_{i=1}^{15} x_i$ is the total of interviews, the sample size. Without loss of generality, we choose to not use the category "undecided", then fourteen frequencies are used, $x_1, \ldots, x_{14}$, $n = \sum_{i=1}^{14} x_i$. Taking the candidates ordering we have $x_1$ being the vote frequency of Alvaro Dias, $x_2$ being the vote frequency of Cabo Daciolo, ..., $x_{13}$ being the vote frequency of Vera Lucia, and $x_{14}$ being the frequency of blank votes. From now on, we use only the respective number of each candidate, instead of his(her) name.

We associate to each candidate a ball with the candidate number, and our aim is to estimate the proportions of balls with the different numbers in the population. In this way, the multinomial model is a natural construction for the result of the polls, and the model parameters, $\theta_1, \ldots, \theta_{14}$, are the proportion of votes, in the population, for each class, the candidates and the blank votes. The multinomial model is

$$\Pr(X_1 = x_1, \ldots, X_{14} = x_{14} | \theta_1, \ldots, \theta_{14}) = \frac{n!}{\prod_{i=1}^{14} x_i!} \prod_{i=1}^{14} \theta_i^{x_i},$$

for which $\sum_{i=1}^{14} \theta_i = 1, 0 \leq \theta_i \leq 1, i = 1, \ldots, 14$.

The Dirichlet prior and posterior densities are respectively the functions

$$f(\theta_1, \ldots, \theta_{14} | a_1, \ldots, a_{14}) = \frac{\Gamma(\sum_{i=1}^{14} a_i)}{\prod_{i=1}^{14} \Gamma(a_i)} \prod_{i=1}^{14} \theta_i^{a_i - 1} \text{ and}$$

$$f(\theta_1, \ldots, \theta_{14} | a_1, \ldots, a_{14}, x_1, \ldots, x_{14}) = \frac{\Gamma(\sum_{i=1}^{14} a_i + x_i)}{\prod_{i=1}^{14} \Gamma(a_i + x_i)} \prod_{i=1}^{14} \theta_i^{a_i + x_i - 1},$$

then the prior density is from a Dirichlet distribution with parameter vector $(a_1, \ldots, a_{14})$ and the posterior density is from a Dirichlet with parameter vector $(a_1 + x_1, \ldots, a_{14} + x_{14})$. For simplicity, we will write $(\theta_1, \ldots, \theta_{14} | data)$ to refer to the posterior density of the Dirichlet distribution.

Let us to evaluate the probability of the $i$-th candidate be elected. That is, (s)he can be elected in the first round or must go to the second round and be elected. For a candidate been elected in the first round, it is necessary to achieve more than 50% of the valid votes. In Brazil elections, the blank votes are not considered "valid votes", then for the $i$-th candidate be elected in the first round, it is necessary that $\theta_i > (1 - \theta_{14})/2$. In the second round, the candidate that receive more votes is elected.

Without loss of generality, consider that we are interest in the candidate $i = 1$. The calculus for any other candidate follows the same steps. The probability of candidate 1 be elected is

$$\Pr(candidate\ 1\ is\ elected\ in\ the\ 1^{st}\ round\ |\ data) +$$
$$\{\Pr(no\ candidate\ elected\ in\ the\ 1^{st}\ round)$$
$$\times [\Pr(1\ is\ one\ of\ the\ two\ most\ voted\ in\ the\ 1^{st}\ round\ |\ data)$$
$$\times \Pr(1\ elected\ in\ the\ 2^{nd}\ round\ |\ data)]\}.$$

In other words,

$$\Pr(\theta_1 > (1 - \theta_{14})/2\ |\ 1^{st}\ round) + \left[\left(1 - \sum_{i=1}^{13} \Pr(\theta_i > (1 - \theta_{14})/2\ |\ 1^{st}\ round)\right) \times\right.$$

$$\left.\sum_{j=2}^{13} \left(\Pr(\min(\theta_1, \theta_j) > \max(\theta_3, \ldots, \theta_{j-1}, \theta_{j+1}, \ldots, \theta_{13}) | 1^{st}\ round) \times \Pr(\theta_1 > \theta_j | 2^{nd}\ round)\right)\right].$$

Clearly, $\Pr(\min(\theta_1, \theta_j) > \max(\theta_3, \ldots, \theta_{j-1}, \theta_{j+1}, \ldots, \theta_{13}) \mid 1^{st}\ round)$, the probability of 1 and $j$ be in the 2nd round, is the most complicated factor to be calculate in the above expression. From the Dirichlet distribution, we can evaluate this probability by

$$\Pr(\min(\theta_1, \theta_j) > max(\theta_3, \ldots, \theta_{j-1}, \theta_{j+1}, \ldots, \theta_{13}) \mid 1^{st}\ round) =$$
$$\int f(\theta_1, \ldots, \theta_{14} | data)\, d\theta_1 \cdots d\theta_{14}.$$

This integral, that does not have an analytical solution, is evaluated over the set

$$\{(\theta_1, \ldots, \theta_{14});\ \min(\theta_1, \theta_j) > max(\theta_3, \ldots, \theta_{j-1}, \theta_{j+1}, \ldots, \theta_{13})\}.$$

The solution for such an integral is generally solved using the Monte Carlo method. Since the distribution of $\theta_1, \ldots, \theta_{14} | data$ is known, one could generate a sample from the posterior distribution and approximate the integral by the proportion of possible samples that respect the integration restriction. However, Monte Carlo method may not have the necessary precision, and then we present a new formulation, that simplifies the problem.

The Dirichlet distribution can be generated by a combination of Gamma independent distributions (Pereira and Stern, 2008; Aitchison, 2003). Consider that, $\gamma_i$ has independent Gamma distribution with parameters $\alpha_i$ (shape) and $\beta$ (rate), $i = 1, \ldots, 14$, then $\theta_i = \gamma_i / \sum_{k=1}^{14} \gamma_k$, implies that $\theta_1, \ldots, \theta_{14}$ has Dirichlet distribution with parameters $\alpha_1, \ldots, \alpha_{14}$. One can rewrite the probability of both, 1 and $j$, be in the 2nd round by

$$\Pr\left[\min\left(\frac{\gamma_1}{\sum_{k=1}^{14} \gamma_k}, \frac{\gamma_j}{\sum_{k=1}^{14} \gamma_k}\right) > \max\left(\frac{\gamma_2}{\sum_{k=1}^{14} \gamma_k}, \ldots, \frac{\gamma_{j-1}}{\sum_{k=1}^{14} \gamma_k}, \frac{\gamma_{j+1}}{\sum_{k=1}^{14} \gamma_k}, \ldots, \frac{\gamma_{13}}{\sum_{k=1}^{14} \gamma_k}\right)\right].$$

As $\sum_{k=1}^{14} \gamma_k$ is a positive random variable that divides all Gamma random variables, this probability can be simplified to

$$\Pr[\min(\gamma_1, \gamma_j) > \max(\gamma_2, \ldots, \gamma_{j-1}, \gamma_{j+1}, \ldots, \gamma_{13})].$$

Consider $\delta_{min} = \min(\gamma_1, \gamma_j)$ and $\delta_{max} = \max(\gamma_2, \ldots, \gamma_{j-1}, \gamma_{j+1}, \ldots, \gamma_{13})$, then we have that the distribution and density functions of $\delta_{min}$ are

$$F_{min}(u) = 1 - [1 - F_{\gamma_1}(u)][1 - F_{\gamma_j}(u)],$$
$$f_{min}(u) = f_{\gamma_1}(u)[1 - F_{\gamma_j}(u)] + f_{\gamma_j}(u)[1 - F_{\gamma_1}(u)].$$

The distribution function of $\delta_{max}$ is

$$F_{max}(u) = \prod_{k=2}^{j-1} F_{\gamma_k}(u) \prod_{k=j+1}^{13} F_{\gamma_k}(u).$$

Also, $\Pr[\min(\gamma_1, \gamma_j) > \max(\gamma_2, \ldots, \gamma_{j-1}, \gamma_{j+1}, \ldots, \gamma_{13})] = \Pr[\delta_{min} > \delta_{max}] =$
$$\int_0^\infty F_{max}(u)\, dF_{min}(u) = \int_0^\infty F_{max}(u)\, f_{min}(u) du,$$

which is a single integration that can be easily computed by any numerical integration method. Note that, $\delta_{sum} = \sum_{k=2}^{13}\gamma_k$ has Gamma distribution with parameters $\sum_{k=2}^{13}\alpha_k$ (shape) and $\beta$ (rate), and using similar ideas, we have that

$$\Pr(\theta_1 > (1-\theta_{14})/2 \mid 1^{st}\ round) = \Pr\left(\gamma_1 > \sum_{k=2}^{13}\gamma_k \mid 1^{st}\ round\right) = \int_0^\infty F_{sum}(u) f_{\gamma_1}(u) du,$$

and

$$\Pr(\theta_1 > \theta_j \mid 2^{nd}\ round) = \Pr(\gamma_1 > \gamma_j \mid 2^{nd}\ round) = \int_0^\infty F_{\gamma_j}(u) f_{\gamma_1}(u) du,$$

where $F_{sum}(\cdot)$ is the distribution function of $\delta_{sum}$.

The great advantage of our solution is that we can compute the probabilities using numerical calculus, for simple functions. It does not depend on generation of random variables as in Monte Carlo methods, then our solution can be considered as an exact solution. There are many different methods to solve integrals (Davis and Rabinowitz, 1984) that give the desired precision in the computations, and are fast and simple to use. We used an adaptive quadrature method (Piessens et al., 1983) that is available in R Software (R Core Team, 2018).

The results obtained for the Brazilian president's election are presented in figures 1 to 4. For both companies' data, the probability of any candidate be elected in the $1^{st}$ round is zero (for all polls' data). Also, the results are indicating that candidates 5 and 9 have now probability 1 to be in the $2^{nd}$ round, for both companies. The main difference in the results is about which candidate will be elected. The data from Datafolha is indicating that candidate 5 has probability 1 to be elected, and the data from IBOPE are indicating that candidates 5 and 9 has almost the same probability to be elected, with a little advantage for candidate 9. These differences may be a consequence of the date of the most recent poll of each company.

In these notes, we present a simple way to compute complicated posterior probabilities in contingency tables. As motivation we used the data of election poll of two Brazilian companies for the Brazilian president election 2018. The presented results are very interesting, and both figures 1 to 4 shows that the Brazilian election can change from a day to another. All election polls used in these notes are for the September month, where the $1^{st}$ round of election will be in October 7, 2018. Our main conclusion who will be the next Brazilian president is unpredictable. However, we may conclude that we have probability one that the $2^{nd}$ round will occur, and the candidates 5 and 9 will be the two most voted in the $1^{st}$ round.

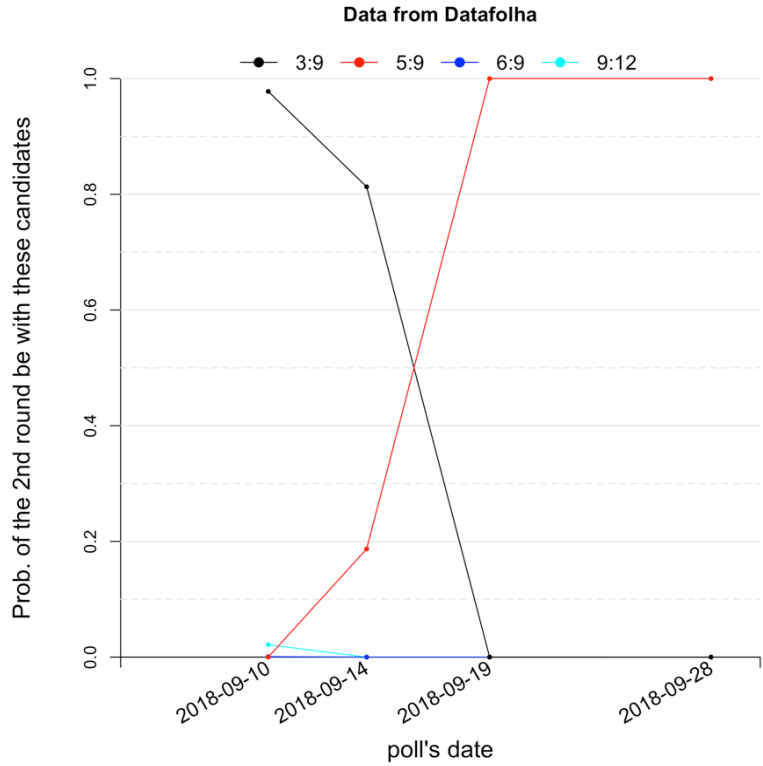

Figure 1. Probability of a specific pair of candidates be the two most voted in the 1st round; the probabilities were evaluated from the polls' data of the company Datafolha.

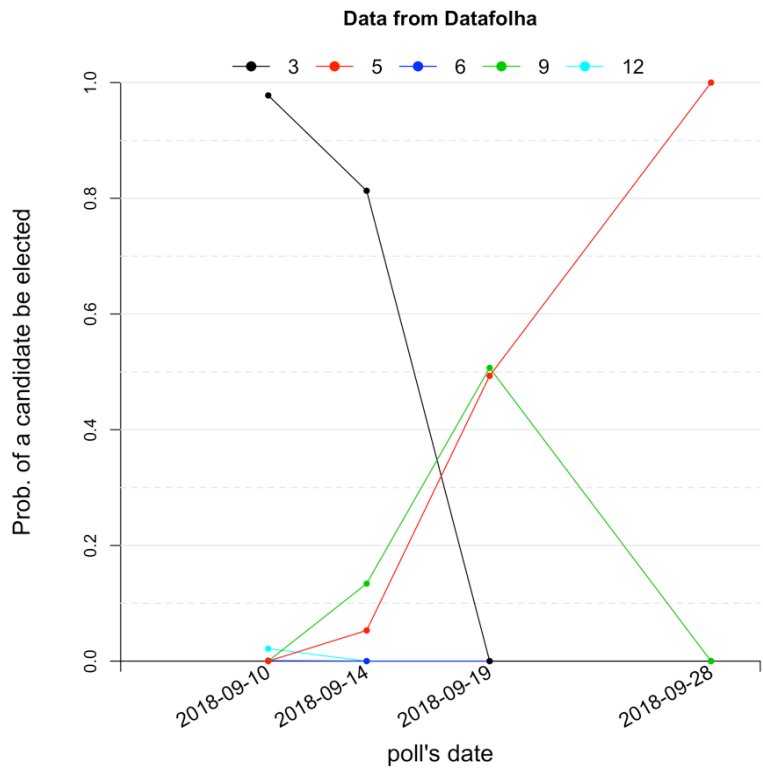

Figure 2. Probability of a specific candidate be elected at the end of election process; the probabilities were evaluated from the polls' data of the company Datafolha.

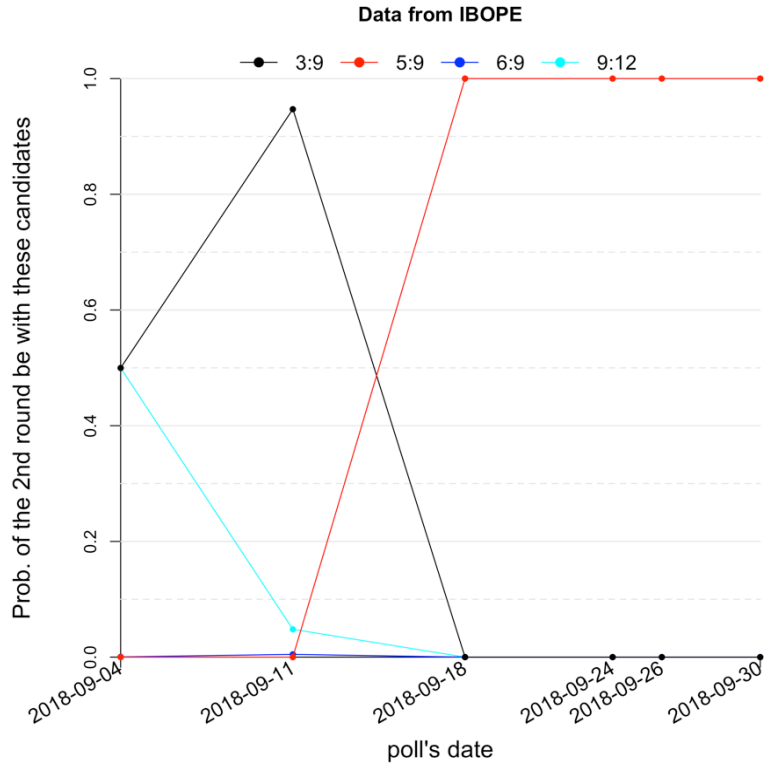

Figure 3. Probability of a specific pair of candidates be the two most voted in the 1st round; the probabilities were evaluated from the polls' data of the company IBOPE.

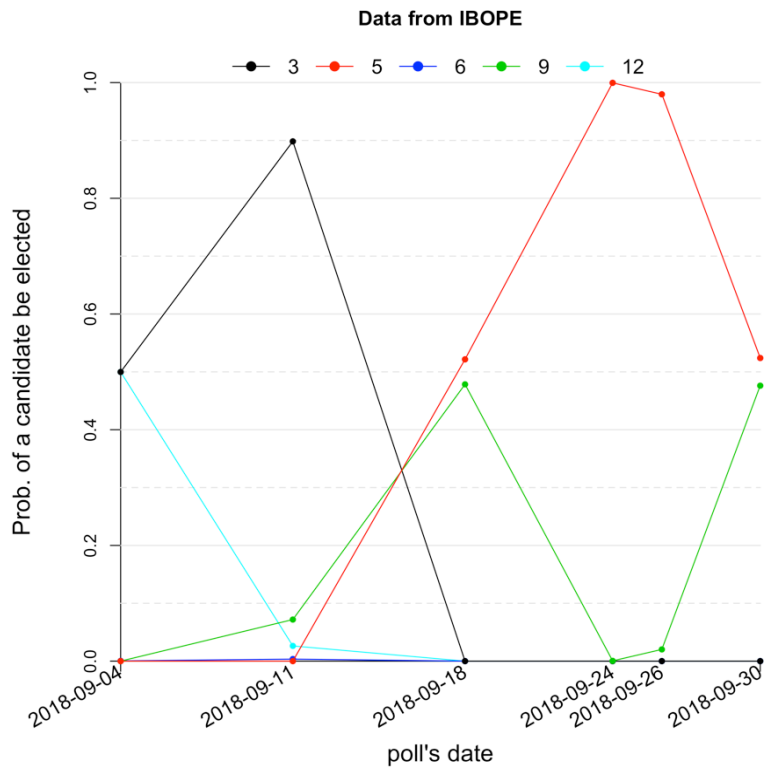

Figure 4. Probability of a specific candidate be elected at the end of election process; the probabilities were evaluated from the polls' data of the company IBOPE.